*Article*

# Development of Charging/Discharging Scheduling Algorithm for Economical and Energy-Efficient Operation of Multi-EV Charging Station

**Hojun Jin [1], Sangkeum Lee [2], Sarvar Hussain Nengroo [1] and Dongsoo Har [1,\*]**

[1] Cho Chun Shik Graduate School of Mobility, Korea Advanced Institute of Science and Technology (KAIST), 291, Daehak-ro, Yuseong-gu, Daejeon 34141, Korea; hjjin1995@kaist.ac.kr (H.J.); sarvar@kaist.ac.kr (S.H.N.)

[2] Environment ICT Research Section, Electronics and Telecommunications Research Institute (ETRI), 218, Gajeong-ro, Yuseong-gu, Daejeon 34129, Korea; sangkeum@etri.re.kr

\* Correspondence: dshar@kaist.ac.kr; Tel.: +82-42-350-1267

**Abstract:** As the number of electric vehicles (EVs) significantly increases, the excessive charging demand of parked EVs in the charging station may incur an instability problem to the electricity network during peak hours. For the charging station to take a microgrid (MG) structure, an economical and energy-efficient power management scheme is required for the power provision of EVs while considering the local load demand of the MG. For these purposes, this study presents the power management scheme of interdependent MG and EV fleets aided by a novel EV charging/discharging scheduling algorithm. In this algorithm, the maximum amount of discharging power from parked EVs is determined based on the difference between local load demand and photovoltaic (PV) power production to alleviate imbalances occurred between them. For the power management of the MG with charging/discharging scheduling of parked EVs in the PV-based charging station, multi-objective optimization is performed to minimize the operating cost and grid dependency. In addition, the proposed scheme maximizes the utilization of EV charging/discharging while satisfying the charging requirements of parked EVs. Moreover, a more economical and energy-efficient PV-based charging station is established using the future trends of local load demand and PV power production predicted by a gated recurrent unit (GRU) network. With the proposed EV charging/discharging scheduling algorithm, the operating cost of PV-based charging station is decreased by 167.71% and 28.85% compared with the EV charging scheduling algorithm and the conventional EV charging/discharging scheduling algorithm, respectively. It is obvious that the economical and energy-efficient operation of PV-based charging station can be accomplished by applying the power management scheme with the proposed EV charging/discharging scheduling strategy.

## 1. Introduction

Recently, energy shortage and atmospheric pollution problems are becoming critical issues, as high energy consumption and carbon emissions are caused by traditional combustion engine vehicles. As an effective way to solve these problems, electric vehicles (EVs) have become an essential component in the transportation sector to reduce dependence on oil resources and improve environmental performance with various types of renewable energy resources in the microgrid (MG). In addition, plug-in electric vehicles (PEVs), plug-in hybrid electric vehicles (PHEVs), and fuel cell vehicles (FCVs) are not only cost-effective but also eco-friendly in terms of energy prices and operating expenses compared to combustion engine vehicles [1]. As EV infrastructure rapidly grows, EVs provide the function of stabilizing the power system. For technological advances in power management of EV infrastructure, an EV aggregator is utilized to offer energy-efficient and cost-effective charging/discharging strategies. With batteries equipped in EVs, they can operate as an energy storage system (ESS) shifting load demand from peak to peak-off hours and reducing electricity costs [2]. A three-step approach is proposed to manage PHEVs' charging load to minimize the cost for energy suppliers [3]. Moreover, EVs are described as controllable loads which can be utilized to stabilize the grid supporting the energy in vehicle-to-building (V2B) or vehicle-to-home (V2H) applications, and can also serve as spinning reserves in specific conditions [4–6].

With the widespread usage of EVs, the structure of the parking station is presented to satisfy the electrical energy requirements of EV owners. These requirements of parking stations are satisfied by interacting with the utility grid. The parking station satisfies the accessibility to the charging station for EVs and provides an opportunity to integrate the EVs' batteries. Moreover, when vehicle-to-grid (V2G) or grid-to-vehicle (G2V) technologies are used for EVs, batteries of EVs at parking stations can serve as a flexible reserve capacity [7]. Due to their ability to store energy, parking stations can participate in the energy and reserve markets. In addition, parking stations can perform a prominent role to supply charging services for EV owners. Generally, conventional EV chargers, such as AC Level 2 (charging rate is 10–22 kW) and direct current quick charging (DCQC) (charging rate is 50–120 kW), are used in public charging stations [8]. As various types of chargers have different charging rates and characteristics, different charging modes can be assigned to EVs depending on the EV charging requirement. As more chargers are installed in a charging station, more installation and operating costs are incurred. Consequently, it is significant to use a limited number of chargers efficiently. Several studies were presented to satisfy multiple EVs' charging requirements, such as the development of intelligent charging strategies [9,10], and optimization of charging stations' location and size [11,12]. Cooperative charging strategies will lead to an increase in operational efficiency and the reduction in electricity costs.

EV charging stations generally produce high charging load in the distribution network when added to the local power systems. Several studies have been implemented, which are relevant to quantifying the variation of peak load [13,14]. With an uncoordinated charging scheme for PHEVs located in the metropolitan distribution network of Australia, peak load shifting was required to make the stable operation of the distribution network [15]. In [16], the impact of uncontrolled EV charging on local power distribution networks was analyzed, and the distribution networks were stabilized by incorporating a coordinated charging scheme. Moreover, the optimal strategy, which controls the charging activities in the parking station, was presented by analyzing the effect of EV charging on daily load demand [17]. In [18], a two-stage demand response model is described to coordinate the peak load incurred by charging EVs. To reduce the stress on the grid, renewable energy systems (RESs) can be installed in the charging station infrastructure and utilized with smart charging strategies to charge EVs. Additionally, the photovoltaic (PV)-based EV charging station can participate in the energy markets or EV arbitrage markets as the grid-connected PV systems are widely deployed and EV markets rapidly grow. To maximize the revenues of the PV-based EV charging station and minimize the battery energy storage system (BESS) capacity fading, a multi-objective optimization is applied [19]. A dynamic searching peak and valley algorithm is applied to PV-based EV charging stations to mitigate the effect on the public grid and reduce the electricity cost of the public grid. In [20], the impact of a PV-based charging station on the economics and emissions from the power grid was analyzed. A study in [21] described smart control strategies to integrate EVs and PV systems with the future electricity systems. A PV generation can be used to supplement increased peak electricity demands due to the mid-day charging of PHEVs [22]. In [23], the effect of fast EV chargers on a retail building's load demand was analyzed, and 38% of the PHEV charging load was supplied with demand management and produced PV power.

V2G technologies can be applied to power distribution systems of larger capacity, facilitate a smart grid with power management, and provide ancillary services to the users [24]. Through interactions between smart grid and EV aggregator in the V2G applications, charging–discharging coordination can be achieved with supply-demand equilibrium as V2G operates faster than traditional power plants, eliminating the problems such as stress on the power grid, power balance, urgent charging demand, and energy deviation [25,26]. Moreover, V2G provides backup capabilities for renewable energy resources, such as wind and solar power generations, efficiently supplementing intermittent power production of RESs [27,28]. According to the total number of EVs in a local network, distributed power capacity provided by V2G can have a profound effect on the power system [29]. In [30], the particle swarm optimization (PSO)-based approach was presented to improve balances between operating cost and emission in the V2G framework. V2G has the effect of reducing the overall cost of services and the electricity bills of customers and improving load factors by selling energy to the utility grid.

With the utilization of V2G, intelligent charging–discharging methods are carried out in real-time when the grid system requires an economical and energy-efficient solution and ancillary services. Intelligent charging–discharging methods can be applied to the grid system where EVs and charging station are interconnected to a data communication network, and the charging station is linked to a transmission-distribution system operator. It can also track, control, and constrain the use of electric devices in a charging station to optimize the electricity demand. Optimization techniques are usually used for energy management of the grid system, and the optimized operation of the power system is determined based on optimization variables, such as charging power, charging–discharging status, price prediction, and power balance [31–33]. With the cooperative charging–discharging operations of EVs, the vehicle-to-vehicle (V2V) charging mechanism can be utilized to stabilize the electricity network, which is beneficial to both EVs having the role of energy consumers and energy providers [34]. A semi-distributed online V2V charging strategy based on electricity price control was presented to obtain high revenue for discharging and low cost for charging EVs [35]. When traded with the electric company, the commercial price of electricity produced from the RESs is generally determined based on the system marginal price (SMP) and the renewable energy certificates (RECs) issued by authorities such as the Korea Power Exchange (KPX) [36]. In study [37], a real-time model performing V2G power transactions is applied for EV fleets. The electrical energy can be sold/purchased to/from the utility grid through bi-directional charging in V2G networks [38,39].

In the charging station, a communication system is significant to extract each EV's operation, such as charging and discharging, and its charging rate [40]. Smooth and reliable communication is an essential component to operate charging stations with sharing networks and effectively scheduling EVs' operations for the customers [41]. To prevent compatibility problems between charging stations, a diversity of internet-based communication methods have been presented and communication standards have been established [42,43]. Data and energy flow of charging stations are bidirectionally transmitted among EVs, charging stations, and the utility grid. The ISO/IEC 15118 standard is defined as a complementary international standard supporting internet protocol-based bidirectional communication system [44]. It is mainly used to develop improved and autonomous charging/discharging control mechanism between EVs and charging station. The IEC 61850 is used to standardize EV charging systems consisting of AC or DC and define requirements of general EV, electric vehicle supply equipment (EVSE), and different charging modes. Additionally, its communication standard defines information models to be applied to the information exchange between electrical devices in the charging station [40].

Electric power industry has significant impact on our daily life, and stable and efficient electricity supply can maintain the balanced operation of the power system [45]. Therefore, accurately forecasting the change of load demand is essential to ensure the stable power system and reliable electricity market. In addition, as RESs are integrated with MGs, they are utilized to supply electricity to households, buildings, factories, and EV charging stations. Among various types of RESs, many studies are currently implemented for PV output forecasting due to its characteristic of intermittent power generation. PV power forecasting is helpful for planning and scheduling of power distribution, energy management, increasing the operational efficiency of the power system, and minimizing the electricity cost. Typically, short-term forecasting of electricity load demand and RESs' output can be used for power system control or energy flow scheduling of electrical components in the MG.

Recently, deep learning (DL) is applied to improve the accuracy of data forecasting since a large amount of data can be collected from several sensing devices. Different DL algorithms, such as convolutional neural network (CNN), recurrent neural network (RNN), long short-term memory (LSTM), gated recurrent unit (GRU), and other hybrid deep neural network architectures, are generally used as beneficial tools in time-series data forecasting [46]. The function of DL based neural network is to learn complex mappings from the patterns of input and output datasets by supporting numerous inputs and outputs. The advantage of artificial neural network (ANN) includes that the solution is very fast and simple compared with the physics-based models. In addition, it does not require knowledge about parameters on internal system, and it can solve the complex mathematical problems with high accuracy. RNN is generally used to deal with time-series data, however, it is easy to have long-term dependence. To prevent long-term dependence on RNN, the gate structure of LSTM performs the function of selective memory of past information. After LSTM, the GRU network is also proposed with another RNN gate structure, which has only two gates used to control the influence of datasets in previous time, to solve problems of RNN such as vanishing or exploding gradient. GRU has high forecasting accuracy and alleviates the overfitting problem of LSTM [47]. In [48], LSTM based framework is applied to forecast a residential load by capturing temporal power consumption pattern in single-meter load profile. The LSTM-RNN was presented to establish an accurate PV power forecasting model reflecting the temporal changes in PV power production [49]. The GRU neural network is used to establish the model to forecast the electricity load with stacked auto-encoding compressing the historical dataset [50]. The hybrid DL approach was developed based on the sequence-to-sequence autoencoder and GRU for short-term PV power prediction, which has higher performance than other DL models [51].

In this paper, multi-EV charging/discharging scheduling strategy is proposed for the power management of interdependent MG and EV fleets located in a charging station. The proposed charging/discharging scheduling algorithm is presented to solve the problem of the grid system instability that may occur when the charging demand of EVs is added to the local load demand, and to minimize the operating cost of the PV-based charging station while satisfying the charging requirements of parked EVs. A portion of produced PV power and discharging power of EVs can be supplied to the local load demand and sold to the electric company at the sum of SMP and weighted REC, and the rest of these power can be directly supplied to the charging demand of EVs. If the PV power and discharging power of EVs are insufficient for the load demand, the power is supplied from the utility grid. It is assumed that the charging station is located in workplace, and EVs can be charged or discharged during the parking period that is determined by the distribution of arrival time at home and workplace. The proposed power management scheme is mainly composed of two stages: prediction and scheduling. Using a time series model trained by the GRU network, the local load demand and PV power production are forecasted in the prediction stage. In the scheduling stage, parked EVs are assigned to the charging/discharging operations depending on the EV charging/discharging scheduling algorithm considering the future trend of the local load demand and PV power output. For the coordinated power distribution of power sources and EV charging/discharging scheduling, multi-objective optimization is applied to the MG and EV fleets to minimize the operating cost of the PV-based charging station and grid dependency of the MG and maximize the utilization of EVs as ESSs while considering each EV's charging requirements. The contributions of this paper are listed as follows.

(1) A novel power management scheme for independent MG and EV fleets in the PV-based charging station is presented. The PV power produced from the PV-based charging station is used as a secondary source to satisfy the local load demand and charging demand of EVs, and it is consumed prior to grid power. With the proposed EV charging/discharging scheduling algorithm, discharging power from parked EVs is also be used to meet the local load demand and charging demand of EVs. In addition, the system operator of the PV-based charging station can decrease the operating cost by selling the produced PV power and discharging power of EVs to electric companies which are supplemented to the local load demand. For economical and energy-efficient power management of the PV-based charging station, a demand response (DR) program, SMP, and REC are considered.

(2) The multi-EV charging/discharging scheduling algorithm with GRU network is proposed for economical and energy-efficient power management of parked EVs in the PV-based charging station. Parked EVs of the PV-based charging station are assigned to charging/discharging operation based on the maximum amount of discharging power of EVs determined by the future trends of local load demand and PV power production that are predicted by the GRU network. Compared with the power management scheme with conventional EV charging/discharging scheduling algorithm utilizing only the currently measured local load demand and PV power production, the power management scheme with the proposed EV charging/discharging scheduling algorithm considering the variation trends of local load demand and PV power production can further reduce the operating cost of the PV-based charging station. The proposed power management scheme is also compared with the conventional scheme using the EV charging scheduling algorithm not considering discharging of EVs. As a result of comparing the performance of the power management schemes presented in this paper, a substantial difference is observed in the aspect of operational efficiency and economic feasibility of the PV-based charging station

(3) In the PV-based charging station, several EVs are connected to the charging system via individual EV charging connectors. Since the parked EVs are interconnected to the MG through the electricity network, some EVs can be charged by the utility grid, PV system, and discharging of other EVs, and some EVs can be discharged to supplement the local load demand and charging demand of EVs among EV charging/discharging candidates. In this study, driving pattern of EVs is considered according to distribution of arrival at home and workplace. The EV charging/discharging candidates are established based on each EV's information such as predetermined parking time including margin time, current state of charge (SOC) value, and initiation of charging/discharging process. Using the proposed EV charging/discharging scheduling strategy, parked EVs of the PV-based charging station are utilized as ESSs in the MG.

(4) In the proposed power management scheme, multi-objective optimization is applied for the power management of interdependent MG and EV fleets, which determines the coordinated power distribution of power sources and the charging/discharging operation of parked EVs. Multi-objective optimization for optimal power management of nanogrid is presented in [52]. Through the multi-objective optimization, the operating cost of the PV-based charging station and grid dependency of the MG is minimized, and the utilization of EV charging/discharging is maximized for economical and energy-efficient operation of the PV-based charging station. In addition, the collected information such as remaining parking time and the current SOC of EVs are considered to satisfy the charging requirements of parked EVs, and the operating SOC range and the initiation of charging/discharging process are reflected to prevent over-charging and over-discharging and improve battery health.

This paper is organized as follows. In Section 2, the system architecture including the PV-based charging station is described. Section 3 explains the proposed power management scheme for obtaining the charging/discharging scheduling of EVs in the PV-based charging station. Section 4 shows the simulation results of the comparison of power management schemes, and Section 5 presents the conclusion of this study.

## 2. System Model Formulation

### 2.1. Overall System Architecture

In this paper, the power management scheme of interdependent MG and EV fleets is presented to schedule charging/discharging operations of parked EVs in PV-based charging station considering the local load demand and PV power production. In the grid-connected mode, the utility grid is linked to the electricity network through the point of common coupling (PCC), and the electricity load is also connected to the electricity network to be supplied power from power resources, such as the utility grid, PV system, and discharging of EVs. For data communication in the MG, the microgrid central controller (MGCC) is used as communication interface to maintain the energy balance of the electricity network. Typically, as the number of EVs accommodated in the charging station increases, the electricity network can be quite unstable due to the increasing peak load caused by electricity load of the charging station. Consequently, power management is essential for the charging station to manage EVs participation and charging/discharging scheduling.

For the economical and energy-efficient EV charging/discharging scheduling of the PV-based charging station, the system operator of the charging station needs to consider various information such as the state of grid system, weather information, electricity purchasing/selling price, and EV information. The system operator receives the information of the local load demand in the grid system and electricity selling prices such as SMP and REC from the KPX. Weather information is obtained from the Korea Meteorological Administration (KMA), and maximum PV power output is calculated by solar irradiance data in weather information. The electricity rate purchased from the utility grid is determined based on the DR program specified by the Korea Electric Power Corporation (KEPCO). The charging station receives specific information relevant to charging/discharging conditions of EVs such as current SOC, target SOC, arrival time, and scheduled departure time.

Charging batteries is a critical task to manage proper operation of sensor network [53], transportation system [54], and many more. In the PV-based charging station, EVs can be utilized as a power supply/demand component while parked. When an EV is parked and connected to the charging system, the system operator receives the parked EVs' information. Based on this information, a charging/discharging operation is applied to each EV in real-time with the proposed EV charging/discharging scheduling algorithm considering the local load demand and generated PV power output. The PV-based charging station operator is responsible for charging each parked EV to target SOC within a predetermined parking time while maintaining the stable electricity network. The frequent charging and discharging cycle of EV incurs the reduction in the usable lifetime of the EV battery. Due to the battery degradation, the EV owner's permission is necessary to utilize the EV to the V2G application. For high penetration of EVs in V2G system, an economic effect needs to be provided to the EV owners who permits to participate in V2G service as well as charging station operator. In the charging station, a parking fee discount can be served as an economic benefit for EV owners.

The overall system architecture of the PV-based charging station is presented in Figure 1. The utility grid is linked to the electricity network via transformer, and electricity can be supplied to the local load demand and charging demand of EVs through the electricity network. The PV system installed on the charging station is connected to the AC bus with a DC-AC converter, and the produced PV power can be supplied to the local load demand and charging demand of EVs according to the power management scheme. The parked EVs are interconnected to the electricity network using bidirectional charger that is a state-of-the-art EV charger capable of charging and discharging power from an EV battery. The EVs having a bidirectional charging capability can be utilized to give electricity to load demand, feed energy back into the utility grid, and provide backup electricity in case of a power outage or emergency. In this system, the discharging power of EVs, which is determined depending on the EV charging/discharging scheduling algorithm, can also be used to supplement the local load demand and charging demand of EVs. When the produced PV power and discharging power of EVs are insufficient for total load demand, the grid power is used. The key role of the proposed EV charging/discharging scheduling algorithm is to decrease the peak load that incurred by the simultaneous charging demand of EVs and minimize the operating cost of the PV-based charging station by selling the PV power and discharging power of EVs to the electric market while supplying them to local load demand.

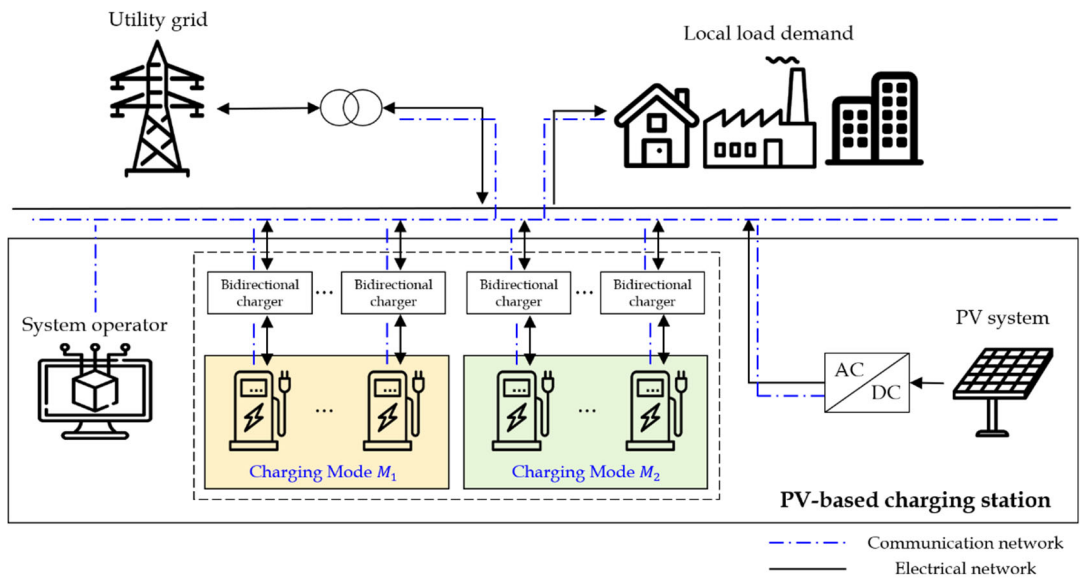

**Figure 1.** Overall system architecture of the PV-based charging station.

In the PV-based charging station, three charging modes, such as $M_0$ (0 kW), $M_1$ (7 kW), and $M_2$ (19.2 kW), are utilized to satisfy EV owners with different charging requirements and improve the charging service quality [55]. The $M_1$ and $M_2$ mode are included in the level 2 charger which is described as a primary charging method typically used in public facilities [56]. Depending on the EVs' information, such as the predetermined parking time, the current SOC, and target SOC, the charging mode is assigned to each EV. The determination of charging mode for each EV is shown as follows.

$$CM_i = \begin{cases} M_1 & T_{p,i} \geq T_{c,i}^{M_1} + T_w \\ M_2 & T_{c,i}^{M_1} + T_w > T_{p,i} \geq T_{c,i}^{M_2} + T_w \qquad \forall i \in I \\ M_0 & otherwise \end{cases} \tag{1}$$

where the identification number of an EV which requires charging is represented by the variable $i \in 1,2,3,\cdots,I$, the charging mode of $i$-th EV is represented by $CM_i$, the predetermined parking time of $i$-th EV is represented by $T_{p,i}$, the minimum required charging time of $i$-th EV in $M_1$ and $M_2$ mode is represented by $T_{c,i}^{M_1}$ and $T_{c,i}^{M_2}$, and the margin time considering discharging of EV is represented by $T_w$. $T_{c,i}^{M_1}$ and $T_{c,i}^{M_2}$ are calculated based on the assumption that each EV is only charged without discharging in the PV-based charging station. In order to utilize EV discharging for the power management of the MG, $T_w$ is added to $T_{c,i}^{M_1}$ and $T_{c,i}^{M_2}$, respectively, to determine the charging mode of each EV. For $i$-th EV, if $T_{p,i}$ is larger than the sum of $T_{c,i}^{M_1}$ and $T_w$, the charging mode is assigned to $M_1$ mode. If $i$-th EV is not assigned to $M_1$ mode and the $T_{p,i}$ is larger than the sum of $T_{c,i}^{M_2}$ and $T_w$, the charging mode is assigned to $M_2$ mode. If $i$-th EV is not assigned to $M_1$ or $M_2$ mode, the EV is denied entry to the charging station and its charging mode is assigned to $M_0$ mode. After assigning the charging mode to parked EVs, on-off strategy is applied for EV charging/discharging as it is more practical method than regulating the charging rate to control a number of EVs.

To describe the driving pattern of a vehicle, the distribution of arrival time at home and workplace is used as shown in Figure 2. The commuter's daily moving pattern during weekdays is considered for the distribution of arrival time. The driving pattern is developed from [57], after analyzing the parking time and driving time of commuters in [58]. From the characteristics of the routine trip of commuters, it is observed that more than 60% of vehicles arrive at workplace from 7:20 to 13:20 h and at home from 15:00 to 21:00 h. As a time interval is considered to be 15 min, the total number of time slots for a day becomes 96 slots. Then all vehicles can be divided into several groups depending on these arrival times. If the daily schedule is predetermined, the commuter can establish the operation of V2G and G2V, and location. In this paper, it is assumed that the PV-based charging station is located at the workplace. The parking periods of EVs in the workplace are decided depending on the duration between arrival time at workplace and arrival time at home. The duration in which the EV is parked in the workplace is treated as the slot for charging/discharging of the EV. Considering the discharging of EVs, the charging/discharging process is performed on EVs having more than 10 h of predetermined parking time. Moreover, each EV is assumed to have 64 kWh capacity of battery and the target SOC value of each EV is set to 80% while parked. The initial SOC value of each EV is provided depending on the normal distribution with the mean value as 15 and the standard deviation value as 5.

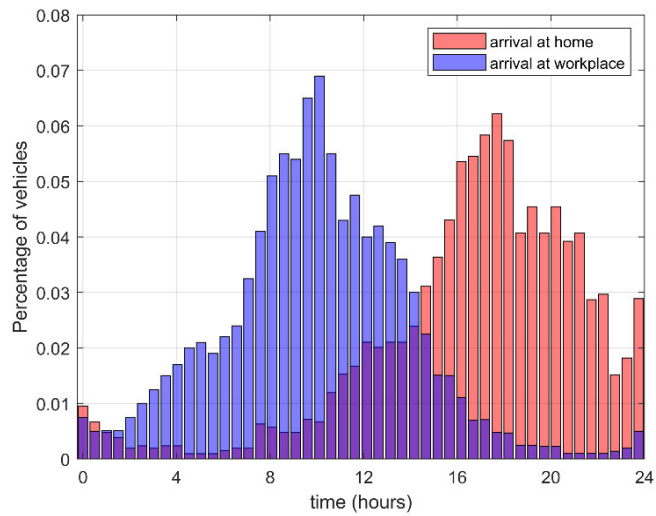

**Figure 2.** Distribution of arrival time of vehicles at home and workplace.

## 2.2. Energy Supply/Demand Model of the PV-based Charging Station

### 2.2.1. EV Model

$$\sum_{i=1}^{I} PW_{ch}^i \times O_{ch}^i(t) = PW_{grid,EV}(t) \times O_{grid,EV}(t) + PW_{PV,EV}(t) \times O_{PV,EV}(t) + PW_{EV,EV}(t) \times O_{EV,EV}(t), \qquad \forall t \qquad (2)$$

$$\sum_{i=1}^{I} PW_{dch}^i \times O_{dch}^i(t) = PW_{EV,load}(t) \times O_{EV,load}(t) + PW_{EV,EV}(t) \times O_{EV,EV}(t), \qquad \forall t \qquad (3)$$

Charging and discharging demand of EVs are described in (2) and (3), where $PW_{ch}^i$ and $PW_{dch}^i$ denote the charging/discharging power of $i$-th EV in assigned charging mode, $O_{ch}^i(t)$ and $O_{dch}^i(t)$ denote the switch function of charging and discharging of $i$-th EV at time $t$, $PW_{grid,EV}(t)$, $PW_{PV,EV}(t)$, and $PW_{EV,EV}(t)$ denote the grid power, the PV power, and the discharging power of parked EVs supplied to charging demand of EVs at time $t$, $O_{grid,EV}(t)$, $O_{PV,EV}(t)$, and $O_{EV,EV}(t)$ denote the switching functions of $PW_{grid,EV}(t)$, $PW_{PV,EV}(t)$, and $PW_{EV,EV}(t)$, $PW_{EV,load}(t)$ denotes the discharging power of parked EVs supplied to the local load demand at time $t$, $O_{EV,load}(t)$ denotes the switching function of $PW_{EV,load}(t)$. $\sum_{i=1}^{I} PW_{ch}^i \times O_{ch}^i(t)$ and $\sum_{i=1}^{I} PW_{dch}^i \times O_{dch}^i(t)$ represent the total charging and discharging demand of parked EVs at time $t$, respectively. In the PV-based charging station, the power from the utility grid, the PV system, and discharging of some EVs are supplied to meet the charging demand of the rest of EVs. The discharging power from parked EVs is used to supplement the local load demand and the charging demand.

### 2.2.2. PV Power Model

$$PW_{PV}(t) = PW_{PV,load}(t) \times O_{PV,load}(t) + PW_{PV,EV}(t) \times O_{PV,EV}(t), \qquad \forall t \qquad (4)$$

The PV power, generated from the PV system installed in the charging station, is described in (4), where $PW_{PV}(t)$ is the total amount of produced PV power at time $t$, $PW_{PV,load}(t)$ is the PV power supplied to the local load demand at time $t$, and $O_{PV,load}(t)$ is the switching function of $PW_{PV,load}(t)$. A part of PV power is directly used for the charging demand of EVs from the PV system, and the rest of PV power is supplemented for the local load demand.

### 2.2.3. Grid Power Model

$$PW_{grid}(t) = PW_{grid,p}(t) - PW_{s,grid}(t), \qquad \forall t \qquad (5)$$

$$PW_{grid,p}(t) = PW_{grid,load}(t) \times O_{grid,load}(t) + PW_{grid,EV}(t) \times O_{grid,EV}(t), \qquad \forall t \qquad (6)$$

$$PW_{s,grid}(t) = PW_{PV,load}(t) \times O_{PV,load}(t) + PW_{EV,load}(t) \times O_{EV,load}(t), \qquad \forall t \qquad (7)$$

The grid power consumption of the overall system architecture can be expressed by (5)–(7), where $PW_{grid}(t)$ is the total grid power consumption at time $t$, $PW_{grid,p}(t)$ is the power purchased from the grid at time $t$, $PW_{s,grid}(t)$ is the power sold to the grid at time $t$, $PW_{grid,load}(t)$ is the grid power supplied to the local load demand at time $t$, and $O_{grid,load}(t)$ is the switching function of $PW_{grid,load}(t)$. When the produced PV power and the discharging power of EVs are insufficient for local load demand, the power is purchased from the grid to meet the demand. In (7), the PV power and the discharging power of EVs are sold to the grid and used to satisfy the local load demand.

### 2.2.4. Local Load Demand Model

$$PW_{load}(t) = PW_{grid,load}(t) \times O_{grid,load}(t) + PW_{PV,load}(t) \times O_{PV,load}(t) + PW_{EV,load}(t) \times O_{EV,load}(t), \forall t \qquad (8)$$

The local load demand model is described by (8), where $PW_{load}(t)$ is the local load demand except the charging demand of EVs at time $t$. To satisfy the local load demand, represented by $PW_{load}(t)$, the power can be supplied from the utility grid, PV system, and discharging of EVs.

### 2.3. Electricity Rate (DR program, SMP, and REC)

The DR program is typically utilized to establish an efficient and flexible power system by providing an incentive to users to consume less electricity when electricity rate or overall electricity consumption is high [59]. In the PV-based charging station, the charging demand of EVs can be scheduled to reduce the peak load and the electricity cost based on the DR program. According to the DR program, the electricity rate is presented as shown in Figure 3, where $0.055/kWh during 23:00–09:00, $0.108 kWh during 09:00–10:00, 12:00–13:00, and 17:00–23:00, and $0.179/kWh during 10:00–12:00 and 13:00–17:00. The Renewable Portfolio Standard (RPS) is applied as renewable energy policy to decrease economic burden of Electric Power Industry Foundation and activate investment of renewable energy business to power providers. The renewable energy provider can obtain a REC, which describes a tradable commodity, by generating and offering 1 MWh of electricity from renewable energy sources. In addition, energy providers can get additional revenue by selling REC to the energy market. To increase the amount of investment in renewables, weighted value is provided to RECs by government depending on the type, technology, and size of facility. Renewable energy providers having high technology or small-scale facilities cab obtain benefits from this policy of weighted REC system. In this paper, weight for the PV system installed on the PV-based charging station is provided as 1.2, as the PV system is installed on general sites and its capacity is less than 100 kW [60]. The SMP is described as the maximum average cost among scheduled electric energy generators. In the energy market, the price of renewable energy is set by the sum of SMP and weighted REC depending on the fixed price contract. Figure 4 and Figure 5 show monthly average value of the local load demand of the MG and the PV power production of PV-based charging station, respectively. The power management scheme of the MG with EV charging/discharging scheduling algorithm is performed using multi-objective optimization based on the comparison of the local load demand and PV power production.

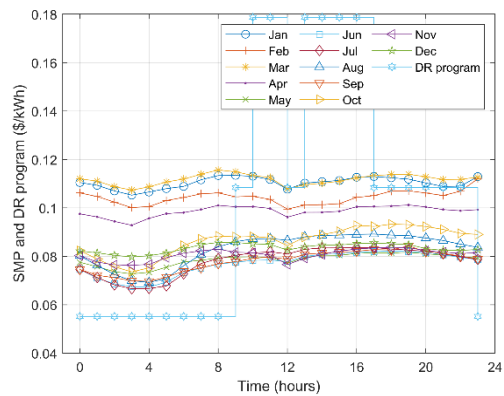

**Figure 3.** Electricity cost depending on average monthly SMP and DR program.

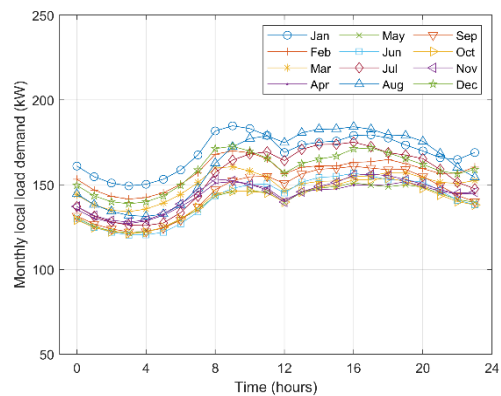

**Figure 4.** Local load demand of the MG.

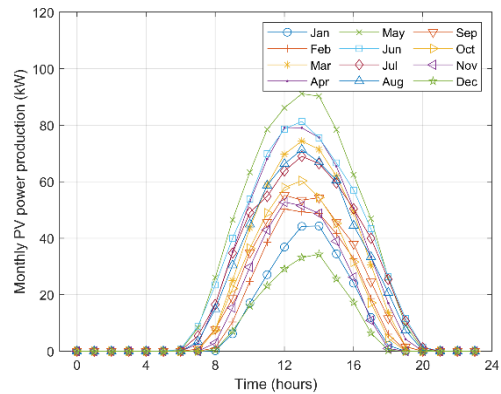

**Figure 5.** PV power production.

### 2.4. Prediction of Local Load Demand and PV Power Production

In the system architecture, parked EVs in the PV-based charging station are utilized as ESS to maintain stable and efficient power system. As shown in Figure 1, the local load demand, the PV system, and the parked EVs are connected to the electrical network, and they can communicate mutually through the communication network. Depending on the forecasted data of the local load demand and the PV power output, the charging/discharging operation of EVs is determined by the system operator, and it can decrease the peak load of the entire system and increase the economic feasibility of the PV-based charging station. The PV power and the scheduled discharging power of EVs can be sold to electric company while providing electricity to the local load demand. When the power is insufficient to satisfy the charging demand and the local load demand, the necessary power is supplied from the utility grid.

RNN is a kind of ANN that can process time-series data or sequential data, memorizing long term dependencies [61]. However, as time lags increase, the vanishing/exploding gradient problem may occur in the training process of RNN. The structure of LSTM and GRU were proposed to resolve the above-mentioned problem. The GRU is presented to obtain better performance in the aspect of computational time and cost, compared with the LSTM. In this work, the GRU network is utilized to predict the variation trends of the local load demand and PV power output. As shown in Figure 6a, the applied GRU network is mainly composed of two parts: 6 GRU layers and 3 fully connected layers. The appropriate structure of GRU network is established by trial-and-error process [62]. The structure of GRU cell is presented in Figure 6b. In the GRU cell, there exist two control gates, namely update gate ($z_t$) and reset gate ($r_t$). The update gate regulates how much the hidden state information at prior time step ($h_{t-1}$) will be delivered to the current time step $t$. The more update gate is activated, the more hidden status information at the previous time step is reflected. The reset gate controls how much the hidden state information at prior time step will be discarded before delivered to the current time step $t$. The smaller the reset gate, the more information at prior time step is ignored [63].

To construct the forecast model, the GRU networks are trained with the 1-year dataset of the local load demand and PV power production, respectively. The dataset of the local load demand is obtained from the KPX, and the dataset of the PV power output generated from the PV panel is calculated based on the following components, such as solar irradiances and ambient temperatures, provided from the KMA [64]. Each dataset is split into a training set (80%) and a validation set (20%). The GRU network is trained with each dataset, in other words, the first network is trained with local load demand dataset, the second with PV power output dataset. To reflect the variational trend of the local load demand and the PV power output with power management scheme, the numbers of inputs and outputs are set as 12 and 7, respectively. In the GRU network, $PW(t-m)$, $PW(t-m+1)$, …, and $PW(t)$ are the inputs of local load demand or PV power output at $t$-th time interval where $m = 11$, and $PW(t + 1)$, $PW(t + 2)$, ..., and $PW(t + n)$ can be the outputs of local load demand or PV power output at $t$-th time interval where $n = 7$. The performance of the forecast model for the local load demand and the PV power output is evaluated by root-mean-squared-error (RMSE). The RMSE is defined as $\sqrt{\frac{1}{N}\sum_{i=1}^{N}(\hat{X}_i - X_i)^2}$, where $\hat{X}_i$ represents the predicted value of local load demand and PV power output, and $X_i$ represents the real value of local load demand and PV power output. The RMSE of the forecast model is 3.8 and 6.74 for each validation dataset.

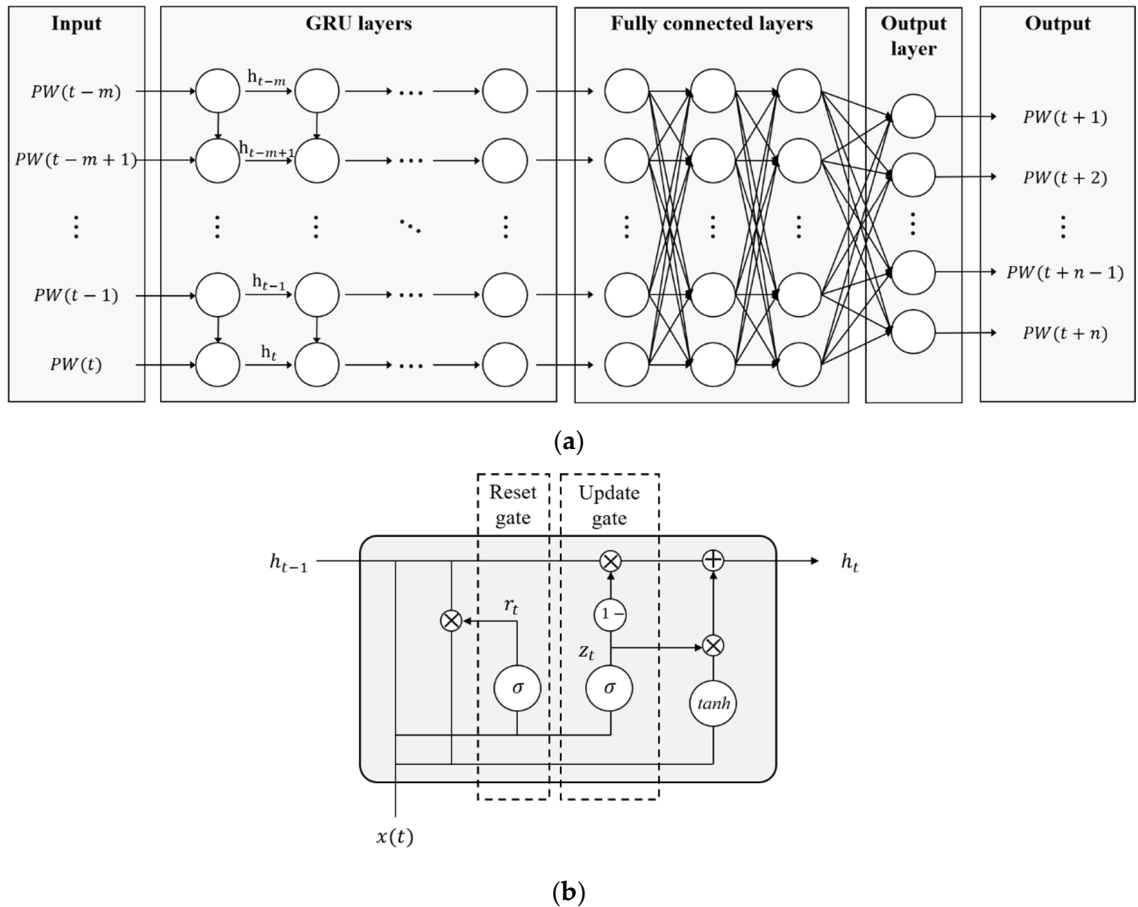

**(a)**

**(b)**

**Figure 6.** Prediction of local load demand and PV power production: (**a**) GRU network architecture; (**b**) GRU cell.

## 3. Power Management Scheme of MG with EV Charging/Discharging Scheduling Algorithm

In this section, features of the power management scheme of the MG with the proposed EV charging/discharging scheduling algorithm is presented. It is assumed that each EV can be charged or discharged through bidirectional EV charger linked with the PV-based charging station. The bidirectional EV charger can control the switching of the connected EV between charging and discharging operations. Maximum amount of the discharging power of the parked EVs is provided depending on the variational trends of the local load demand and PV power production. According to the EV's information, each parked EV becomes a candidate for charging/discharging operation. Then, by performing the multi-objective optimization, the coordinated power distribution of power sources is carried out for the MG and a "charging", "discharging", or "idle" operation is assigned to the switching function of each parked EV based on the EV charging/discharging candidates. The switching function of each EV is applied to the multi-objective optimization considering the operating cost of PV-based charging station, PV power consumption, grid dependency, utilization of EVs, current SOC value of EVs, and the remaining parking time of EVs simultaneously.

### 3.1. Proposed EV Charging/Discharging Scheduling Algorithm

In the conventional EV charging/discharging scheduling algorithm, maximum amount of the discharging power from parked EVs is given depending on the specific condition relevant to the difference between local load demand and PV power production, as presented in (9). The determined maximum amount of discharging power is considered for the EV charging/discharging scheduling in the power management scheme using multi-objective optimization.

$$\begin{cases} 0 \leq \sum_{i=1}^{I} PW_{dch}^i \times O_{dch}^i(t) < PW^{dch,\max(1)} & if\ PW_{load}(t) - PW_{PV}(t) \geq PW^{flag} \\ 0 \leq \sum_{i=1}^{I} PW_{dch}^i \times O_{dch}^i(t) < PW^{dch,\max(2)} & if\ PW_{load}(t) - PW_{PV}(t) < PW^{flag} \end{cases} \tag{9}$$

where $PW^{flag}$ is the flag parameter to decide the period that discharging of EVs is more or less activated, and $PW^{dch,max(1)}$ and $PW^{dch,max(2)}$ are the maximum amount of discharging power from parked EVs in each circumstance. $PW^{dch,max(1)}$ has higher value than $PW^{dch,max(2)}$, which means discharging of EVs can be more utilized to the power management of the MG with $PW^{dch,max(1)}$ than with $PW^{dch,max(2)}$. If the difference between local load demand and PV power output at the $t$-th time interval $PW_{load}(t) - PW_{PV}(t)$ is equal to or bigger than $PW^{flag}$, the maximum amount of discharging power of EVs becomes $PW^{dch,max(1)}$. On the contrary, if it is smaller than $PW^{flag}$, the maximum amount of discharging power of EVs becomes $PW^{dch,max(2)}$. In the proposed EV charging/discharging scheduling algorithm, the maximum amount of discharging power of EVs is determined depending on the following conditions.

$$\begin{cases} 0 \leq \sum_{i=1}^{I} PW_{dch}^i \times O_{dch}^i(t) < PW^{dch,max(1)} & if\ \sum_{k=0}^{K} \left( PW_{load}(t+k) - PW_{PV}(t+k) \right) \geq (K+1) \times PW^{flag} \\ 0 \leq \sum_{i=1}^{I} PW_{dch}^i \times O_{dch}^i(t) < PW^{dch,max(2)} & if\ \sum_{k=0}^{K} \left( PW_{load}(t+k) - PW_{PV}(t+k) \right) < (K+1) \times PW^{flag} \end{cases} \tag{10}$$

For the power management of the MG, the current and future states of local load demand and PV power production are used to determine the maximum amount of discharging power of EVs. By considering the future trends of local load demand and PV power production, a more appropriate charging/discharging operation can be allocated to each parked EV in a specific circumstance. The future states of local load demand and PV power production are forecasted by the GRU network, trained with each training dataset, as presented in Section 2. If the cumulative sum of difference between local load demand and PV power production from the time interval $t$ to $t + K$ $\sum_{k=0}^{K} \left( PW_{load}(t+k) - PW_{PV}(t+k) \right)$ is equal to or bigger than $(K+1) \times PW^{flag}$, the maximum amount of discharging power from parked EVs becomes $PW^{dis,max(1)}$. On the contrary, if it is smaller than $(K+1) \times PW^{flag}$, the maximum value of discharging power becomes $PW^{dis,max(2)}$.

### 3.2. Multi-Objective Optimization for Power Management of Interdependent MG and EV fleets

As the maximum amount of discharging power of EVs is given, the EV candidates to be charged or discharged are determined depending on the information of each EV, such as the remaining parking time, the current SOC, and initiation of charging/discharging process at $t$-th time interval. The EV can belong to the candidate to be charged if the initiation of charging process is less than the maximum initiation and the current SOC value is less than the maximum SOC value. On the contrary, the EV can belong to the candidate to be discharged if the initiation of discharging process is less than the maximum initiation and the current SOC value is higher than the minimum SOC value. However, the EV cannot belong to the candidate to be discharged if the remaining parking time is less than or equal to the required time to charge up to the target SOC. Then, the multi-objective optimization is applied to determine the functioning behavior of EVs among charging, discharging, and idle operations for the power management of the MG. If the EV is included in the candidate to be charged, the charging or idle operation can be applied to the EV, and if the EV is included in the candidate to be discharged, the discharging or idle operation can be applied to the EV. The details of the multi-objective optimization are as follows.

### 3.2.1. Constraints

While the EV in the PV-based charging station is interconnected to the electricity network through the bidirectional EV charger having DC-to-AC inverter, the EV can be charged by the power supplied from the utility grid and, PV system installed on the charging station, and discharging of EVs, and can be discharged to stabilize the power network and decrease the operating cost of PV-based charging station by using the discharging power for the local load demand and charging demand of EVs.

$$PW_{EV}^i(t) = \eta_{ch} \times PW_{ch}^i \times O_{ch}^i(t) - \frac{1}{\eta_{dch}} \times PW_{dch} \times O_{dch}^i(t), \quad \forall i, t \tag{11}$$

$$SOC_{EV}^i(t) = SOC_{EV}^i(t-1) + PW_{EV}^i(t), \quad \forall i, t \tag{12}$$

The charging/discharging power and the SOC value of *i*-th EV are presented in (11) and (12), where $PW_{EV}^i(t)$ represents the charging/discharging power of *i*-th EV at time *t*, $\eta_{ch}$ and $\eta_{dch}$ represent the converter efficiency of the battery of EVs in charging/discharging process, and $SOC_{EV}^i(t)$ represents the SOC value of *i*-th EV at time *t*. The dynamic adjustment of SOC of *i*-th EV can be calculated as shown in (12).

$$SOC_{min}^i \leq SOC_{EV}^i(t) \leq SOC_{max}^i, \quad \forall i, t \tag{13}$$

where $SOC_{min}^i$ and $SOC_{max}^i$ are the minimum and maximum SOC value of the *i*-th EV for all time intervals, respectively. To maintain the efficient operation and prevent the aging of EV battery, the operating range of SOC is set as 20%~80%, as presented in (13). The EV battery can be protected from overcharging and deep discharging issues.

$$O_{ch}^i(t) - O_{ch}^i(t-1) = D_{ch}^{i,+}(t) - D_{ch}^{i,-}(t), \quad \forall i, t \tag{14}$$

$$O_{dch}^i(t) - O_{dch}^i(t-1) = D_{dch}^{i,+}(t) - D_{dch}^{i,-}(t), \quad \forall i, t \tag{15}$$

$$\sum_{t=1}^{T} D_{ch}^{i,+}(t) + D_{ch}^{i,-}(t) \leq N^{max}, \quad \forall i, t \tag{16}$$

$$\sum_{t=1}^{T} D_{dch}^{i,+}(t) + D_{dch}^{i,-}(t) \leq N^{max}, \quad \forall i, t \tag{17}$$

where $D_{ch}^{i,+}(t)$ and $D_{ch}^{i,-}(t)$ represent the positive and negative difference between the previous and the current state of binary variable $O_{ch}^i(t)$ {0,1}, $D_{dch}^{i,+}(t)$ and $D_{dch}^{i,-}(t)$ represent the positive and negative difference between the previous and the current state of binary variable $O_{dch}^i(t)$ {0,1}, and $N^{max}$ represents the maximum initiations of EV's charging/discharging process. In order to decrease the adverse effect of intermittent charging/discharging cycles of EV battery which may incur the reduction in battery capacity and its lifetime [65], the initiations of charging/discharging process of the EV is constrained up to $N^{max}$.

In the PV-based charging station, PV panels with a capacity of about 90 kW is installed on the charging station, and a DC-AC converter is used to supply power to the electricity network via a PV system. For the economical and energy-efficient operation of the PV-based charging station, the produced PV power is used for the local load demand and charging demand of EVs within the maximum PV power generation, as presented in (18), where $PW_{PV}^{max}(t)$ represents the maximum amount of PV power at time $t$.

$$0 \leq PW_{PV,load}(t) \times O_{PV,load}(t) + PW_{PV,EV}(t) \times O_{PV,EV}(t) \leq PW_{PV}^{max}(t), \qquad \forall t \tag{18}$$

In this study, total electricity load demand of the MG is mainly composed of the local load demand and charging demand of EVs, and the electricity is supplied from the utility grid, PV system, and discharging of EVs. In order to prevent excessive power consumption from the utility grid, the following equation is presented.

$$PW_{load}(t) + \sum_{i=1}^{I} PW_{ch}^i \times O_{ch}^i(t) - \left( PW_{PV,load}(t) \times O_{PV,load}(t) + PW_{PV,EV}(t) \times O_{PV,EV}(t) + \sum_{i=1}^{I} PW_{dch}^i \times O_{dch}^i(t) \right)$$

$$\leq PW^{max}, \qquad \forall t \tag{19}$$

where $PW^{max}$ describes the maximum grid power consumption. The difference between the total load demand and the supplied power from the PV system and discharging of EVs represents the grid power consumption. To satisfy the constraint of the grid power consumption, the charging/discharging process of EVs is controlled by (19).

### 3.2.2. Objective Functions

In this paper, minimization of the operating cost of PV-based charging station by (20), minimization of the grid power consumption of the MG by (21), and maximization of the utilization of EVs as ESS by (24) are performed through multi-objective optimization. The charging requirements of parked EVs are also satisfied with the consideration of remaining parking time and current SOC by (22) and (23).

The first objective function is presented to reduce the operating cost of the PV-based charging station, while satisfying the charging requirements of parked EVs. The reduction in operating cost is achieved by the economical and energy-efficient use of grid power, PV power, and discharging power of parked EVs as follows.

$$min \left[ \begin{array}{l} TOU(t) \times \left( \sum_{i=1}^{I} PW_{ch}^i \times O_{ch}^i(t) - PW_{PV,EV}(t) \times O_{PV,EV}(t) - PW_{EV,EV}(t) \times O_{EV,EV}(t) \right) \\ -(SMP(t) + w * REC) \times \left( PW_{PV,load}(t) \times O_{PV,load}(t) + PW_{EV,load}(t) \times O_{EV,load}(t) \right) \end{array} \right] \tag{20}$$

where $TOU(t)$, $SMP(t)$, and $REC$ represent time-of-use (TOU) price which is one of the price-based tariffs under DR program at time $t$, the SMP of electricity supplied from PV power generation and discharging power of parked EVs at time $t$, and REC which can be sold to other electricity company with weighted value $w$, respectively. For the stable and economical operation of the PV-based charging station, the operating cost should be minimized in relation with the EV charging/discharging scheduling by controlling switching functions $O_{ch}^1(t) \cdots O_{ch}^I(t)$, $O_{dch}^1(t) \cdots O_{dch}^I(t)$ of EVs. In the PV-based charging station, after the power required for the charging demand of EVs is self-supplied by the PV system and discharging of EVs, the power from the utility grid is supplied to the rest of the charging demand of EVs at TOU price. The power not used for the charging demand of EVs can be sold to electric company at the price considering the SMP and weighted REC, while using the power to supplement the local load demand.

Due to increasing electricity load demand in the MG, the RES becomes more significant as it has relatively less impact on the environmental issues than non-renewable energy sources. Even though the use of grid power can be more economical according to DR program, the use of PV power is promoted due to its eco-friendly characteristic [66]. Moreover, it can reduce the grid power consumption and enhance energy efficiency by using PV power directly supplied from the PV system, not via long transmission line. The decrement of the grid power consumption and the increment of the PV power consumption can be achieved by the equation as following

$$min\left[\begin{array}{c}\left(PW_{grid,load}(t) \times O_{grid,load}(t) + PW_{grid,EV}(t) \times O_{grid,EV}(t)\right) \\ -\left(PW_{PV,load}(t) \times O_{PV,load}(t) + PW_{PV,EV}(t) \times O_{PV,EV}(t)\right)\end{array}\right] \qquad (21)$$

To satisfy the charging requirements of EVs and utilize the discharging of EVs to the power management of the MG simultaneously, intelligent charging/discharging scheduling process of EVs is essential. After the parked EVs are classified into charging/discharging candidates depending on each EV's information, one of charging, discharging, and idle operations is assigned to each EV by (22) and (23) relevant to the remaining parking time of EVs and the difference between the current SOC and the target SOC of EVs.

$$min\left[\sum_{i=1}^{I} T_{re}^i(t) \times O_{ch}^i(t) - \sum_{i=1}^{I} T_{re}^i(t) \times O_{dch}^i(t)\right] \qquad (22)$$

$$min\left[-(SOC_{max} - SOC_{EV}^i(t))^2 \times O_{ch}^i(t) + (SOC_{max} - SOC_{EV}^i(t))^2 \times O_{dch}^i(t)\right] \qquad (23)$$

$$min\left[-\left(\sum_{i=1}^{I} PW_{ch}^i \times O_{ch}^i(t) + \sum_{i=1}^{I} PW_{dch}^i \times O_{dch}^i(t)\right)\right] \qquad (24)$$

where $T_{re}^i(t)$ represents the remaining parking time of the $i$-th EV before the scheduled departure from the charging station at time $t$. The priority of charging/discharging EV can be provided using $T_{re}^i(t)$ and $SOC_{max} - SOC_{EV}^i(t)$ in (22) and (23). Depending on these equations, if an EV has less remaining parking time and a wide gap between the current SOC and the target SOC, the probability that the EV is assigned to charging operation among EV charging candidates gets higher. On the contrary, if an EV has larger remaining parking time and a narrow gap between the current SOC and the target SOC, the probability that the EV is assigned to discharging operation among EV discharging candidates gets higher. The charging/discharging of parked EVs can be activated and utilized to the power management of the MG according to (24). With these objective functions and constraints mentioned above, the multi-objective optimization is performed by the genetic algorithm (GA) that has advantage of global search ability with its random nature and flexibility [67]. In the multi-objective optimization, the parameters of GA take crossover probability 0.8, mutation probability 0.01, the maximum number of generations 100, and population size 500.

## 4. Simulation Results

### 4.1. Simulation Setup

To verify the effect of the power management scheme with the proposed EV charging/discharging scheduling algorithm, the simulation is executed based on the system architecture described in Figure 1. For the local load and charging demand of EVs, the utility grid, PV system and discharging power of EVs are used as power sources. The proposed EV charging/discharging scheduling algorithm is applied to the power management of the MG to achieve the cost-effective and energy-efficient operation of the PV-based charging station considering the local load demand of the MG. In the simulation, the DR program determines the electricity rate of grid power, and the sum of SMP and weighted REC is used as the selling price of PV power and discharging power of EVs for electric company. It is assumed that the PV-based charging station is located around workplace and the maximum number of EVs that can be parked at the PV-based charging station is 50. Each EV has a battery capacity of 64 kWh, and the SOC values of parked EVs at the moment connected to the charging port are provided depending on the normal distribution with mean $\mu$=15 and standard deviation $\sigma$=5. The expected parking time is determined based on the distribution of arrival time of vehicles at home and workplace presented in

Figure 2. The charging rate is assigned to each EV, depending on its expected parking time and the current SOC, between $M_1$(7 kW) and $M_2$(19.2 kW). To charge EVs up to target SOC, $PW^{max}$ is set to about 157 kW to meet the local load demand and charging demand of EVs, simultaneously. By comparing the difference between local load demand and PV power production with $PW^{flag}$, the maximum amount of discharging power of EVs at each time interval is determined. In the proposed EV charging/discharging scheduling algorithm, $PW^{dch,max(1)}$ is set to 12 kW, indicating that discharging of EVs is relatively more activated, and $PW^{dch,max(2)}$ is set to 2 kW, pointing out that discharging of EVs is relatively less activated. $PW^{flag}$ is set to 86 kW to determine the appropriate EV discharging period according to the amount of local load demand and PV power production. The local load demand and PV power production data used in this study are presented as in Figure 7.

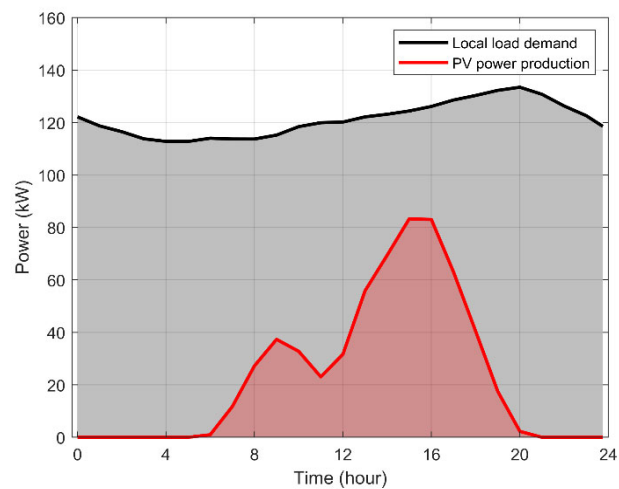

**Figure 7.** Variation of local load demand and PV power production.

In this study, the GRU network is utilized to predict the local load demand of the MG and PV power production of the PV-based charging station. The parameters, adopted from [68] which studied a GRU network with similar structure, are applied to train the GRU network, as presented in Table 1. The total number of training epoch and batch size are set to 1000 and 200 in our simulations. Additionally, the ADAM optimization algorithm is applied with the learning rate of 0.005, gradient moving average of 0.9, dropout rate as 0.2, and gradient threshold as 1 to update weighting coefficients to minimize the error during training the GRU network.

**Table 1.** Training parameters of GRU network.

| | |
|---|---|
| Epoch | 1000 |
| Batch size | 200 |
| Learning rate | 0.005 |
| Gradient moving average | 0.9 |
| Dropout rate | 0.2 |
| Gradient threshold | 1 |

Three different EV charging/discharging scheduling algorithms are applied to the power management scheme of MG in this simulation. The "EV Ch Scheduling" scheme represents the power management scheme of the MG with EV charging scheduling algorithm. The "EV Ch/Dch Scheduling" scheme refers to the power management scheme of MG with conventional EV charging/discharging scheduling algorithm, following the EV discharging condition in (9). The "Proposed EV Ch/Dch Scheduling" scheme represents the power management scheme of MG with the proposed EV charging/discharging scheduling algorithm, following the proposed EV discharging condition in (10). In the Figures 8–13, the green dash-dotted line with square marker, the red dashed line with triangle marker, and the blue solid line with circle marker represent the "EV Ch Scheduling" scheme, the "EV Ch/Dch Scheduling" scheme, and the "Proposed EV Ch/Dch Scheduling" scheme, respectively. At each time interval, the maximum amount of discharging power of EVs, utilized to alleviate peak load and decrease the operating cost of PV-based charging station, is determined by (9) and (10) in the "EV Ch/Dch Scheduling" scheme and the "Proposed EV Ch/Dch Scheduling" scheme, respectively. The EV charging/discharging candidates are determined based on the remaining parking time, the current SOC, and the initiation of charging/discharging process of each EV. Then, through the multi-objective optimization, the coordinated power distribution of power sources in the MG and EVs are allocated to charging/discharging operation among the EV charging/discharging candidates.

### 4.2. Comparative Performance of Power Management Scheme based on the Newly Proposed EV Charging/Discharging Scheduling Algorithm

The variational trends of the total grid power consumption for interdependent MG and EV fleets depending on three EV charging/discharging scheduling strategies, are described in Figure 8. With the "EV Ch Scheduling" scheme, it has lower grid power consumption than the other schemes over the time interval 7:00–9:00, as most of the parked EVs are charged without discharging depending on the EV charging scheduling algorithm. In the "EV Ch/Dch Scheduling" and "Proposed EV Ch/Dch Scheduling" schemes, higher total grid consumption is shown over the time interval 7:00–9:00, because several EVs are discharged for the power management of the MG in the prior time period, and it is available to charge more EVs depending on the relationship between local load demand and PV power production. In addition, higher grid power consumption during 7:00–9:00 causes the reduction in electricity cost imposed for EV charging demand by charging more EVs with low TOU price. Around 13:00, with the "Proposed EV Ch/Dch Scheduling" scheme, less grid power is used for the economic power management of the MG by lessening the number of EVs to be charged due to high-peak periods over the time interval 13:00–16:00, even though the EV charging is preferred depending on the condition $\sum_{k=0}^{K} \left( PW_{load}(t+k) - PW_{PV}(t+k) \right) < (K+1) \times PW^{flag}$. It is observed that the "EV Ch/Dch Scheduling" and "Proposed EV Ch/Dch Scheduling" schemes produce a volatile pattern of total grid power consumption because the power management is deployed to the MG with charging/discharging control of EVs. Due to the relatively higher local load demand than EV charging demand, the different power management schemes of MG with three EV charging/discharging scheduling algorithms have similar variation trends of the total grid power consumption. In addition, the produced PV power is used to reduce the total grid power consumption over the time interval 6:00–20:00.

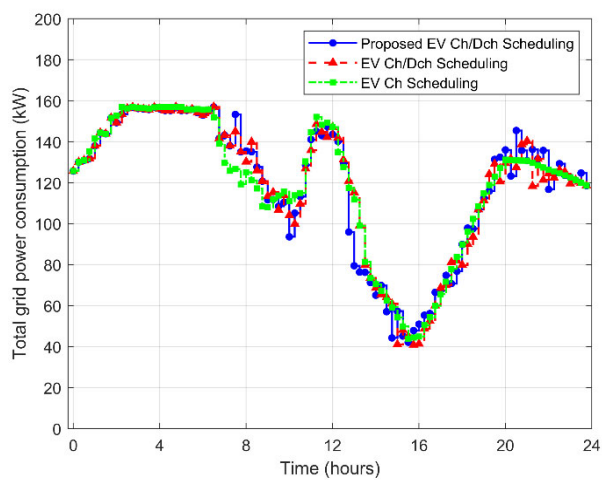

**Figure 8.** Total grid power consumption for interdependent MG and EV fleets.

Figure 9a,b represent the grid power consumption for local load demand and EV charging demand to verify the effect of EV charging/discharging scheduling algorithms on an energy-efficient power usage. With the "EV Ch Scheduling" scheme, higher grid power consumption is shown over the time interval 3:00–7:00, 10:00–12:00, and 19:00–23:00, because the power is supplied to local load demand from the utility grid and PV system without discharging of EVs. With the "EV Ch/Dch Scheduling" and "Proposed EV Ch/Dch Scheduling" schemes, the discharging power of EVs can be used to supply the local load demand and charging demand of EVs. Over the time interval 3:00–7:00, 10:00–12:00, and 19:00–23:00, where discharging of EVs is preferred depending on the condition (9) and (10), the grid power consumption for local load demand decreases due to the discharging power of EVs used for local load demand. In Figure 9b, the "EV Ch Scheduling" scheme has a lower grid power consumption for the charging demand of EVs compared with other schemes utilizing the discharging power of EVs for the power management of the MG, since the discharging of EVs is not used for the charging demand of EVs in the "EV Ch Scheduling" scheme. Moreover, the grid power consumption does not occur over the time interval 17:00–24:00 as the charging of parked EVs is completed by charging them to target SOC in advance, depending on the EV charging scheduling algorithm. With the "EV Ch/Dch Scheduling" and "Proposed EV Ch/Dch Scheduling" schemes, the grid power consumption for the charging demand of EVs has a higher trend over the time interval 3:00–9:00, 11:00–13:00, and 16:00–24:00 as the discharging power of EVs can be used to supplement the charging demand of EVs while considering the power management of the MG to reduce the total grid power consumption. In addition, despite higher local load demand over the time interval 17:00–24:00, the grid power is used for the charging demand to charge EVs that do not reach target SOC due to discharging of EVs.

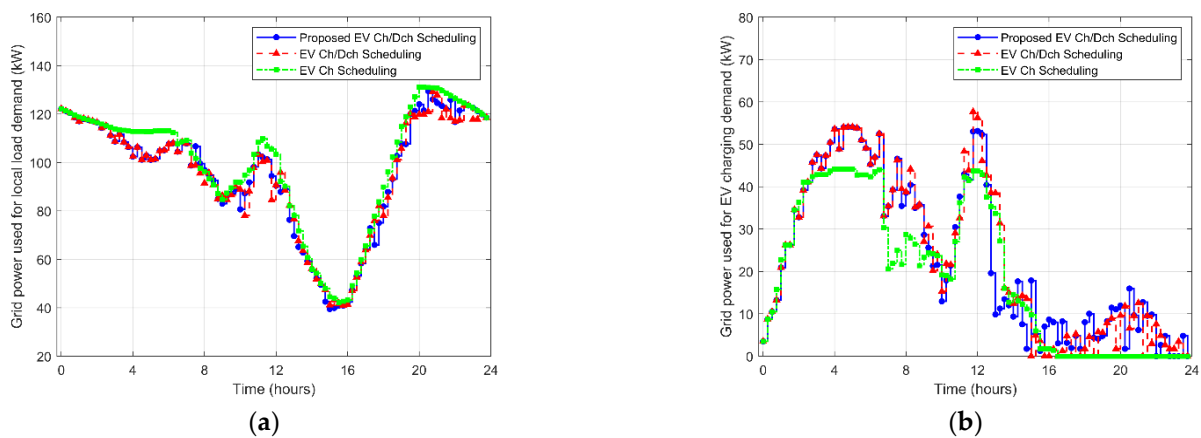

**Figure 9.** (**a**) The grid power consumption for local load demand; (**b**) The grid power consumption for charging demand of EVs.

Figure 10a,b describe the PV power consumption used for local load demand and charging demand of EVs. The PV power produced from PV system installed at the charging station can also be provided to local load demand and charging demand of EVs. In Figure 10a, a high percentage of PV power is used to satisfy the local load demand, showing small difference between three power management schemes. In the aspect of the charging station operator, the system operator can get a revenue by selling the PV power to electric company at the sum of SMP and weighted REC while supplying the PV power to local load demand of the MG. In Figure 10b, a low portion of PV power is consumed for the charging demand of EVs due to the intermittent characteristics of EVs according to the arrival and departure time in the PV-based charging station. With the "Proposed EV Ch/Dch Scheduling" and the "EV Ch/Dch Scheduling" schemes, more PV power is utilized for the charging demand of EVs over the time interval 7:00–9:00 than the "EV Ch Scheduling" scheme, as the charging demand of EVs becomes high due to the discharging of EVs. In addition, by using PV power for charging demand of EVs during 10:00–12:00 corresponding to the high peak periods in the DR program, the operating cost of PV-based charging station can be reduced while reducing the grid power consumption.

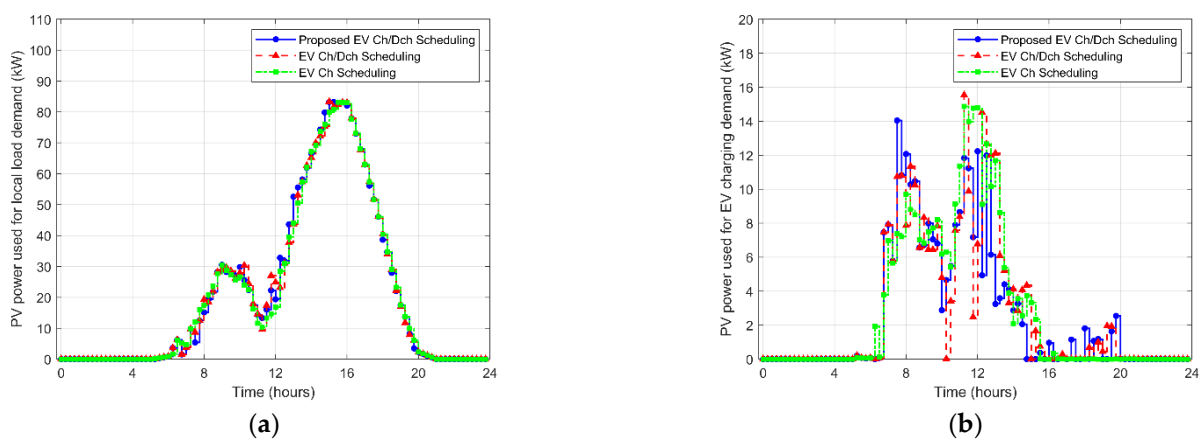

**Figure 10.** (**a**) The PV power consumption for local load demand; (**b**) The PV power consumption for charging demand of EVs.

The charging and the discharging demand of parked EVs in the PV-based charging station are described in Figure 11a,b. Over the time interval 2:00–9:00 and 11:00–13:00, the "EV Ch/Dch Scheduling" and the "Proposed EV Ch/Dch Scheduling" schemes show higher charging demand of EVs than the "EV Ch Scheduling" scheme. It represents that the discharging power from parked EVs can be used for the charging demand of EVs, decreasing the total grid power consumption through the power management of the MG. Around 8:00, the amount of discharging power of EVs that can be used for the power management of the MG with $PW^{max}$ constraint is smaller in the "Proposed EV Ch/Dch Scheduling" scheme than in the "EV Ch/Dch Scheduling" scheme as the "Proposed EV Ch/Dch Scheduling" scheme follows the proposed condition $\sum_{k=0}^{K} \big( PW_{load}(t+k) - PW_{PV}(t+k) \big) < (K+1) \times PW^{flag}$. Furthermore, the "Proposed EV Ch/Dch Scheduling" scheme has less charging demand of EVs than the "EV Ch/Dch Scheduling" scheme according to the decreased amount of discharging power of EVs. With the "Proposed EV Ch/Dch Scheduling" scheme, the charging demand of EVs is reduced over the time interval 12:00–14:00 that has high TOU price by focusing on the EV charging in advance around 8:00. In Figure 11b, it is observed that the maximum amount of discharging power of EVs is limited to 12 kW or 2 kW depending on the conditions (9) and (10). By limiting the total amount of discharging power of EVs at each time interval depending on the difference between local load demand and PV power production, the parked EVs can be charged to the target SOC within the parking time, preventing the excessive discharge of EVs. Through the power management of the MG with the EV charging/discharging scheduling strategy, the discharging power of parked EVs can be used effectively for the local load demand and charging demand of EVs as shown in Figure 12a,b. Due to high charging demand of EVs over the time interval 3:00–8:00, the discharging of EVs is highly activated for the power management of the MG. Furthermore, over the time interval 10:00–12:00 which corresponds to the high peak periods, the discharging power from parked EVs is highly utilized for the local load demand and charging demand of EVs simultaneously depending on the EV charging/discharging scheduling algorithm achieving energy-efficient and economical operation of the PV-based charging station. Though the small amount of discharging power of EVs is available between 12:00 and 17:00, the discharging power of EVs is mainly used for the local load demand to reduce the grid power consumption at high peak periods and decrease the operating cost of the PV-based charging station by selling the power to electric company. With the "Proposed EV Ch/Dch Scheduling" and "EV Ch/Dch Scheduling" schemes, the operating cost of the PV-based charging station can be minimized by charging and discharging multiple EVs over the time interval 18:00–24:00 which has high SMP while satisfying the charging requirements of parked EVs. In Figure 11a, despite the high electricity cost of grid power consumption between 15:00–17:00, larger charging demand of EVs is shown in the "Proposed EV Ch/Dch Scheduling" scheme. With this larger charging demand of EVs in the "Proposed EV Ch/Dch Scheduling" scheme, more discharging power of EVs can be efficiently used to supplement the local load demand and charging demand of EVs over the time interval 17:00–24:00. Depending on the proposed EV discharging condition considering the future trends of local load demand and PV power production, the discharging of parked EVs is utilized earlier in the "Proposed EV Ch/Dch Scheduling" scheme than in the "EV Ch/Dch Scheduling" scheme around 17:00.

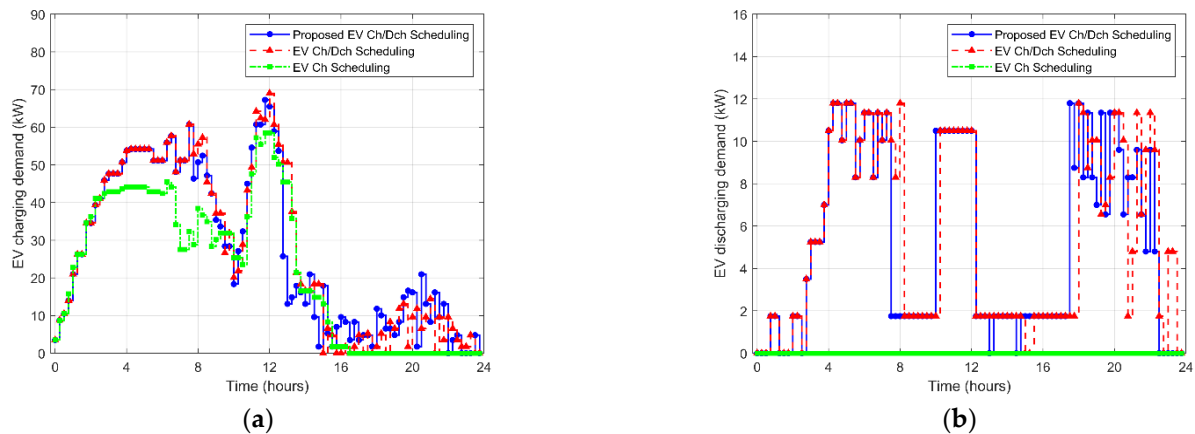

**Figure 11.** (**a**) The charging demand of parked EVs; (**b**) the discharging demand of parked EVs.

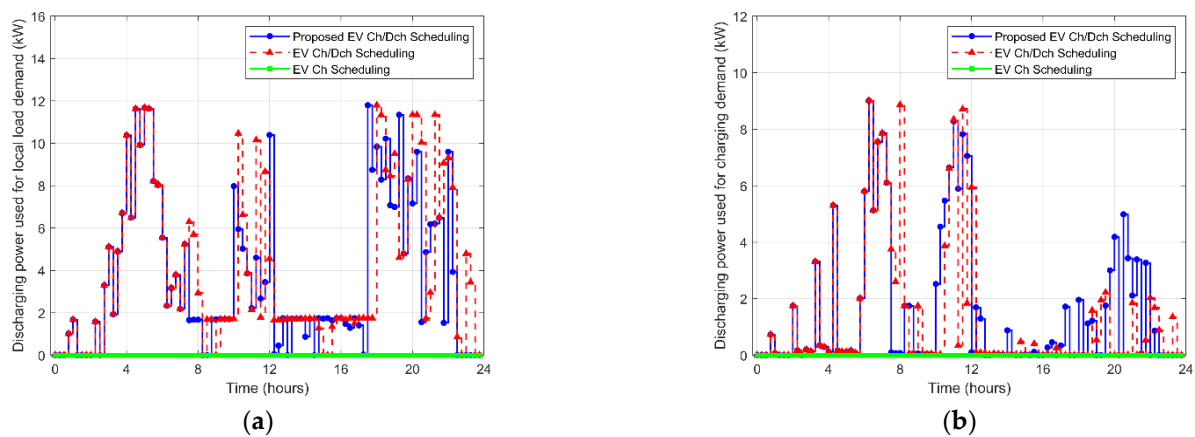

**Figure 12.** (**a**) The discharging power of parked EVs used for local load demand; (**b**) the discharging power of EVs used for charging demand of EVs.

Figure 13a,b describe variation of the sum of remaining parking time of EVs to be charged and discharged, respectively, according to different power management schemes. To utilize parked EVs as ESS for the power management of the MG and satisfy the charging requirements of EVs within predetermined parking time of each EV, (22) is provided in the multi-objective optimization. According to (22), parked EVs with less remaining parking time are likely to be allocated to charging operation, and parked EVs with more remaining parking time are likely to be allocated to discharging operation. With the "EV Ch Scheduling" scheme, a low sum of remaining parking time of parked EVs to be charged is shown over the time interval 2:00–9:00, which represents that a small number of EVs are charged without EVs to be discharged. Compared with the "EV Ch Scheduling" scheme, the "Proposed EV Ch/Dch Scheduling" and "EV Ch/Dch Scheduling" schemes have higher sum of remaining parking time of EVs to be charged because the discharging power from several EVs is directly supplemented to charging demand of EVs. In Figure 13b, it is observed that relatively more EVs are used for power management over the time interval 0:00–8:00, 10:00–12:00, and 18:00–24:00 in which discharging of EVs is highly activated than the other time intervals in which discharging of EVs is less activated. The differences of the sum of remaining parking time of EVs that are discharged between the "Proposed EV Ch/Dch Scheduling" scheme and the "EV Ch/Dch Scheduling" scheme are occurred around 8:00 and 18:00 by considering future trend of local load demand and PV power production in the "Proposed EV Ch/Dch Scheduling" scheme, which helps to perform more economical and energy-efficient operation of the PV-based charging station.

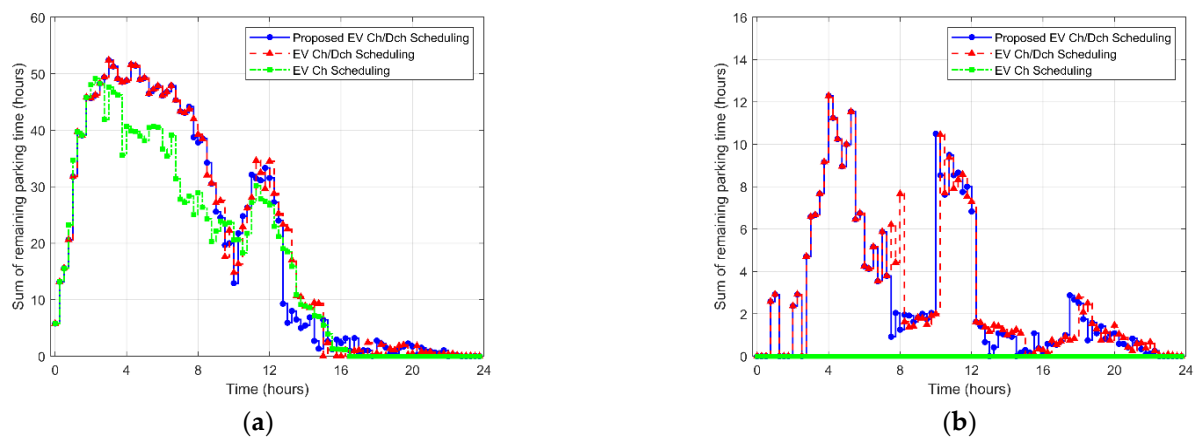

**Figure 13.** (**a**) The sum of remaining parking time of EVs to be charged; (**b**) The sum of remaining parking time of EVs to be discharged.

In Table 2, a summary of the operating cost of the PV-based charging station incurred by different EV charging/discharging scheduling algorithms is described. The power management scheme with the proposed EV charging/discharging scheduling algorithm shows the lowest operating cost of the PV-based charging station, followed by the power management scheme with conventional EV charging/discharging scheduling algorithm. These schemes are followed by the power management scheme with EV charging scheduling algorithm. By using the proposed EV charging/discharging scheduling algorithm, the operating cost of the PV-based charging station decreases by 167.71% and 28.85%, respectively, relative to the power management schemes with EV charging scheduling algorithm and with conventional EV charging/discharging scheduling algorithm. It is notable that the economical and energy-efficient operation of the PV-based charging station can be achieved by performing the coordinated power distribution of power source and utilizing the parked EVs as ESSs depending on the power management scheme of the MG with the proposed EV charging/discharging scheduling algorithm.

**Table 2.** Comparison of operating cost of PV-based charging station according to power scheduling scheme.

| Power Management Scheme | Operating Cost of PV-Based Charging Station |
|:---:|:---:|
| With proposed EV Ch/Dch scheduling | $−35.45 |
| With (conventional) EV Ch/Dch scheduling | $−27.51 |
| With (conventional) EV Ch scheduling | $−13.24 |

## 5. Conclusions

In this study, the power management scheme of interdependent MG and EV fleets aided by a novel EV charging/discharging scheduling strategy is described. To prevent the peak load of the MG that can occur by simultaneous charging demand of EVs and decrease the operating cost of the PV-based charging station, the power management scheme is applied to the MG. The operating cost of PV-based charging station was reduced by the proposed EV charging/discharging scheduling algorithm. For economical and energy-efficient operation of the PV-based charging station, the multi-objective optimization was implemented to minimize the operating cost of the PV-based charging station and grid dependency, and maximize the charging/discharging utilization of EVs, simultaneously. The electricity loads considered in this study were mainly composed of local load demand, which is general electricity load, and the charging demand of parked EVs dispensed from the PV-based charging station. To meet these electricity loads, the electricity is supplied from the utility grid, PV system, and discharging of EVs. For more economical and energy-efficient operation of the PV-based charging station, the future trends of local load demand and PV power production were considered in the proposed EV charging/discharging scheduling algorithm. In the EV charging/discharging scheduling algorithm, the maximum amount of discharging power of EVs is determined based on the difference between local load demand and PV power production, and parked EVs were assigned to charging/discharging operation through multi-objective optimization procedure. The future trends of local load demand and PV power production were predicted by the GRU network. To verify the performance of the proposed EV charging/discharging scheduling strategy, the simulations were performed indicating operating cost reduction, grid dependency reduction, and utilization of EV charging/discharging while satisfying the charging requirements of EVs. With the proposed EV charging/discharging scheduling algorithm, the operating cost of the PV-based charging station is decreased by 167.71% compared with the EV charging scheduling algorithm. The operating cost is decreased by 28.85% relative to the results of the conventional EV charging/discharging scheduling algorithm. From the simulation results, it is evident that the economical and energy-efficient operation of the PV-based charging station can be achieved by actively utilizing parked EVs as ESSs depending on the proposed EV charging/discharging strategy.

**Author Contributions:** Conceptualization, H.J., S.L. and D.H.; methodology, H.J. and S.L.; software, H.J. and S.L.; validation, H.J., S.L. and S.H.N.; formal analysis, H.J.; investigation, H.J.; resources, S.L. and S.H.N.; data curation, S.L. and S.H.N.; writing—original draft preparation, H.J.; writing—review and editing, D.H.; visualization, H.J.; supervision, D.H.; project administration, D.H.; funding acquisition, D.H. All authors have read and agreed to the published version of the manuscript.

**Funding:** This research was funded by the Ministry of Science, ICT, under Grant 2019M3F2A1073314.

**Acknowledgments:** This work was supported by the Energy Cloud Research and Development Program through the National Research Foundation of Korea (NRF) funded by the Ministry of Science, ICT, under Grant 2019M3F2A1073314.

**Conflicts of Interest:** The authors declare no conflict of interest.

## Abbreviations

| | |
|---|---|
| EV | Electric vehicle |
| MG | Microgrid |
| PEV | Plug-in electric vehicle |
| PHEV | Plug-in hybrid electric vehicle |
| FCV | Fuel cell vehicle |
| ESS | Energy storage system |
| V2B | Vehicle-to-building |
| V2H | Vehicle-to-home |
| V2G | Vehicle-to-grid |
| G2V | Grid-to-vehicle |
| DCQC | Direct current quick charging |
| RES | Renewable energy system |
| PV | Photovoltaic |
| BESS | Battery energy storage system |
| PSO | Particle swarm optimization |
| V2V | Vehicle-to-vehicle |
| SMP | System marginal price |
| REC | Renewable energy certificates |
| KPX | Korea Power Exchange |
| EVSE | Electric vehicle supply equipment |
| DL | Deep learning |
| CNN | Convolutional neural network |
| RNN | Recurrent neural network |

| LSTM | Long short-term memory |
|------|------------------------|
| GRU | Gated recurrent unit |
| ANN | Artificial neural network |
| DR | Demand response |
| SOC | State of charge |
| PCC | Point of common coupling |
| MGCC | Microgrid central controller |
| KMA | Korea Meteorological Administration |
| KEPCO | Korea Electric Power Corporation |
| RPS | Renewable Portfolio Standard |
| RMSE | Root-mean-squared-error |
| TOU | Time-of-use |
| GA | Genetic algorithm |

## References

1. Ehsani, M.; Gao, Y.; Longo, S.; Ebrahimi, K.M. *Modern Electric, Hybrid Electric, and Fuel Cell Vehicles*; CRC Press: Boca Raton, FL, USA, 2018.
2. Daina, N.; Sivakumar, A.; Polak, J.W. Electric vehicle charging choices: Modelling and implications for smart charging services. *Transp. Res. Part C Emerg. Technol.* **2017**, *81*, 36–56.
3. Vandael, S.; Claessens, B.; Hommelberg, M.; Holvoet, T.; Deconinck, G. A scalable three-step approach for demand side management of plug-in hybrid vehicles. *IEEE Trans. Smart Grid* **2012**, *4*, 720–728.
4. Dagdougui, H.; Ouammi, A.; Dessaint, L.A. Peak load reduction in a smart building integrating microgrid and V2B-based demand response scheme. *IEEE Syst. J.* **2018**, *13*, 3274–3282.
5. Freund, D.; Lützenberger, M.; Albayrak, S. Costs and gains of smart charging electric vehicles to provide regulation services. *Procedia Comput. Sci.* **2012**, *10*, 846–853.
6. Shin, H.; Baldick, R. Plug-in electric vehicle to home (V2H) operation under a grid outage. *IEEE Trans. Smart Grid* **2016**, *8*, 2032–2041.
7. Parvizimosaed, M.; Farmani, F.; Rahimi-Kian, A.; Monsef, H. A multi-objective optimization for energy management in a renewable micro-grid system: A data mining approach. *J. Renew. Sustain. Energy* **2014**, *6*, 023127.
8. Zhang, Y.; You, P.; Cai, L. Optimal charging scheduling by pricing for EV charging station with dual charging modes. *IEEE Trans. Intell. Transp. Syst.* **2018**, *20*, 3386–3396.
9. Yudovina, E.; Michailidis, G. Socially optimal charging strategies for electric vehicles. *IEEE Trans. Autom. Control* **2014**, *60*, 837–842.
10. Liu, Y.; Deng, R.; Liang, H. Game-theoretic control of PHEV charging with power flow analysis. *AIMS Energy* **2016**, *4*, 379–396.
11. Chung, S.H.; Kwon, C. Multi-period planning for electric car charging station locations: A case of Korean Expressways. *Eur. J. Oper. Res.* **2015**, *242*, 677–687.
12. Lam, A.Y.; Leung, Y.-W.; Chu, X. Electric vehicle charging station placement: Formulation, complexity, and solutions. *IEEE Trans. Smart Grid* **2014**, *5*, 2846–2856.
13. Bandyopadhyay, A.; Leibowicz, B.D.; Beagle, E.A.; Webber, M.E. As one falls, another rises? Residential peak load reduction through electricity rate structures. *Sustain. Cities Soc.* **2020**, *60*, 102191.
14. Dominguez, J.A.; Dante, A.W.; Agbossou, K.; Henao, N.; Campillo, J.; Cardenas, A.; Kelouwani, S. Optimal Charging Scheduling of Electric Vehicles based on Principal Component Analysis and Convex Optimization. In Proceedings of the 2020 IEEE 29th International Symposium on Industrial Electronics (ISIE), Delft, The Netherlands, 17–19 June 2020; pp. 935–940.
15. McCarthy, D.; Wolfs, P. The HV system impacts of large scale electric vehicle deployments in a metropolitan area. In Proceedings of the 2010 20th Australasian Universities Power Engineering Conference, Christchurch, New Zealand, 5–8 December 2010; pp. 1–6.
16. Putrus, G.; Suwanapingkarl, P.; Johnston, D.; Bentley, E.; Narayana, M. Impact of electric vehicles on power distribution networks. In Proceedings of the 2009 IEEE Vehicle Power and Propulsion Conference, Dearborn, MI, USA, 7–10 September 2009; pp. 827–831.
17. Di Silvestre, M.L.; Sanseverino, E.R.; Zizzo, G.; Graditi, G. An optimization approach for efficient management of EV parking lots with batteries recharging facilities. *J. Ambient. Intell. Humaniz. Comput.* **2013**, *4*, 641–649.
18. Hüls, J.; Remke, A. Coordinated charging strategies for plug-in electric vehicles to ensure a robust charging process. In Proceedings of the 10th EAI International Conference on Performance Evaluation Methodologies and Tools on 10th EAI International Conference on Performance Evaluation Methodologies and Tools, Taormina, Italy, 25–28 October 2016; pp. 19–22.
19. Eldeeb, H.H.; Faddel, S.; Mohammed, O.A. Multi-objective optimization technique for the operation of grid tied PV powered EV charging station. *Electr. Power Syst. Res.* **2018**, *164*, 201–211.
20. Tulpule, P.J.; Marano, V.; Yurkovich, S.; Rizzoni, G. Economic and environmental impacts of a PV powered workplace parking garage charging station. *Appl. Energy* **2013**, *108*, 323–332.

21. Zhang, Q.; Tezuka, T.; Ishihara, K.N.; Mclellan, B.C. Integration of PV power into future low-carbon smart electricity systems with EV and HP in Kansai Area, Japan. *Renew. Energy* **2012**, *44*, 99–108.
22. Denholm, P.; Kuss, M.; Margolis, R.M. Co-benefits of large scale plug-in hybrid electric vehicle and solar PV deployment. *J. Power Sources* **2013**, *236*, 350–356.
23. Sehar, F.; Pipattanasomporn, M.; Rahman, S. Demand management to mitigate impacts of plug-in electric vehicle fast charge in buildings with renewables. *Energy* **2017**, *120*, 642–651.
24. Kempton, W.; Tomic, J.; Letendre, S.; Brooks, A.; Lipman, T. *Vehicle-to-Grid Power: Battery, Hybrid, and Fuel Cell Vehicles as Resources for Distributed Electric Power in California*; UC Davis Institute of Transportation Studies: Davis, CA, USA, 2001.
25. Wang, Q.; Liu, X.; Du, J.; Kong, F. Smart charging for electric vehicles: A survey from the algorithmic perspective. *IEEE Commun. Surv. Tutor.* **2016**, *18*, 1500–1517.
26. Nunes, P.; Brito, M. Displacing natural gas with electric vehicles for grid stabilization. *Energy* **2017**, *141*, 87–96.
27. Marano, V.; Rizzoni, G. Energy and economic evaluation of PHEVs and their interaction with renewable energy sources and the power grid. In Proceedings of the 2008 IEEE International Conference on Vehicular Electronics and Safety, Columbus, OH, USA, 22–24 September 2008; pp. 84–89.
28. Kempton, W.; Tomić, J. Vehicle-to-grid power implementation: From stabilizing the grid to supporting large-scale renewable energy. *J. Power Sources* **2005**, *144*, 280–294.
29. Aluisio, B.; Conserva, A.; Dicorato, M.; Forte, G.; Trovato, M. Optimal operation planning of V2G-equipped Microgrid in the presence of EV aggregator. *Electr. Power Syst. Res.* **2017**, *152*, 295–305.
30. Saber, A.Y.; Venayagamoorthy, G.K. Intelligent unit commitment with vehicle-to-grid—A cost-emission optimization. *J. Power Sources* **2010**, *195*, 898–911.
31. Habib, S.; Khan, M.M.; Abbas, F.; Tang, H. Assessment of electric vehicles concerning impacts, charging infrastructure with unidirectional and bidirectional chargers, and power flow comparisons. *Int. J. Energy Res.* **2018**, *42*, 3416–3441.
32. Lambert, G.; Lavoie, S.; Lecourtois, É.; Giumento, A.; Lagace, M.; Dupré, J.-L.; Patault, L.-A.; Boudjerida, N.; Zaghib, K.; Perreault, É. Bidirectional Charging System for Electric Vehicle. Google Patents, 2017.
33. Kaiser, A.; Nguyen, A.; Pham, R.; Granados, M.; Le, H.T. Efficient Interfacing Electric Vehicles with Grid using Bi-directional Smart Inverter. In Proceedings of the 2018 IEEE Transportation Electrification Conference and Expo (ITEC), Long Beach, CA, USA, 13–15 June 2018; pp. 178–182.
34. Zhang, R.; Cheng, X.; Yang, L. Stable matching based cooperative V2V charging mechanism for electric vehicles. In Proceedings of the 2017 IEEE 86th Vehicular Technology Conference (VTC-Fall), Toronto, ON, Canada, 24–27 September 2017; pp. 1–5.
35. Wang, M.; Ismail, M.; Zhang, R.; Shen, X.S.; Serpedin, E.; Qaraqe, K. A semi-distributed V2V fast charging strategy based on price control. In Proceedings of the 2014 IEEE Global Communications Conference, Austin, TX, USA, 8–12 December 2014; pp. 4550–4555.
36. Lee, H.G.; Kim, G.-G.; Bhang, B.G.; Kim, D.K.; Park, N.; Ahn, H.-K. Design algorithm for optimum capacity of ESS connected with PVs under the RPS program. *IEEE Access* **2018**, *6*, 45899–45906.
37. Venayagamoorthy, G.K.; Mitra, P.; Corzine, K.; Huston, C. Real-time modeling of distributed plug-in vehicles for V2G transactions. In Proceedings of the 2009 IEEE Energy Conversion Congress and Exposition, San Jose, CA, USA, 20–24 September 2009; pp. 3937–3941.
38. Garg, S.; Kaur, K.; Ahmed, S.H.; Bradai, A.; Kaddoum, G.; Atiquzzaman, M. MobQoS: Mobility-aware and QoS-driven SDN framework for autonomous vehicles. *IEEE Wirel. Commun.* **2019**, *26*, 12–20.
39. Garg, S.; Kaur, K.; Kaddoum, G.; Ahmed, S.H.; Jayakody, D.N.K. SDN-based secure and privacy-preserving scheme for vehicular networks: A 5G perspective. *IEEE Trans. Veh. Technol.* **2019**, *68*, 8421–8434.
40. Ustun, T.S.; Ozansoy, C.R.; Zayegh, A. Implementing vehicle-to-grid (V2G) technology with IEC 61850-7-420. *IEEE Trans. Smart Grid* **2013**, *4*, 1180–1187.
41. He, Y.; Venkatesh, B.; Guan, L. Optimal scheduling for charging and discharging of electric vehicles. *IEEE Trans. Smart Grid* **2012**, *3*, 1095–1105.
42. Mahmud, K.; Town, G.E.; Morsalin, S.; Hossain, M. Integration of electric vehicles and management in the internet of energy. *Renew. Sustain. Energy Rev.* **2018**, *82*, 4179–4203.
43. Saltanovs, R.; Krivchenkov, A.; Krainyukov, A. Analysis of effective wireless communications for V2G applications and mobile objects. In Proceedings of the 2017 IEEE 58th International Scientific Conference on Power and Electrical Engineering of Riga Technical University (RTUCON), Riga, Latvia, 12–13 October 2017; pp. 1–5.
44. Schmutzler, J.; Wietfeld, C.; Andersen, C.A. Distributed energy resource management for electric vehicles using IEC 61850 and ISO/IEC 15118. In Proceedings of the 2012 IEEE Vehicle Power and Propulsion Conference, Seoul, Korea, 9–12 October 2012; pp. 1457–1462.
45. Carvallo, J.P.; Larsen, P.H.; Sanstad, A.H.; Goldman, C.A. Long term load forecasting accuracy in electric utility integrated resource planning. *Energy Policy* **2018**, *119*, 410–422.
46. Brownlee, J. *Deep learning for time series forecasting: Predict the future with MLPs, CNNs and LSTMs in Python*; Machine Learning Mastery: 2018.
47. Wang, R.; Li, C.; Fu, W.; Tang, G. Deep learning method based on gated recurrent unit and variational mode decomposition for short-term wind power interval prediction. *IEEE Trans. Neural Netw. Learn. Syst.* **2019**, *31*, 3814–3827.

48. Kong, W.; Dong, Z.Y.; Jia, Y.; Hill, D.J.; Xu, Y.; Zhang, Y. Short-term residential load forecasting based on LSTM recurrent neural network. *IEEE Trans. Smart Grid* **2017**, *10*, 841–851.
49. Abdel-Nasser, M.; Mahmoud, K. Accurate photovoltaic power forecasting models using deep LSTM-RNN. *Neural Comput. Appl.* **2019**, *31*, 2727–2740.
50. Ke, K.; Hongbin, S.; Chengkang, Z.; Brown, C. Short-term electrical load forecasting method based on stacked auto-encoding and GRU neural network. *Evol. Intell.* **2019**, *12*, 385–394.
51. Rai, A.; Shrivastava, A.; Jana, K.C. A Robust Auto Encoder-Gated Recurrent Unit (AE-GRU) Based Deep Learning Approach for Short Term Solar Power Forecasting. *Optik* **2022**, *252*, 168515.
52. Lee, S.; Lee, J.; Jung, H.; Cho, J.; Hong, J.; Lee, S.; Har, D. Optimal power management for nanogrids based on technical information of electric appliances. *Energy Build.* **2019**, *191*, 174–186.
53. Moraes, C.; Har, D. Charging distributed sensor nodes exploiting clustering and energy trading. *IEEE Sens. J.* **2016**, *17*, 546–555.
54. Hwang, K.; Cho, J.; Park, J.; Har, D.; Ahn, S. Ferrite position identification system operating with wireless power transfer for intelligent train position detection. *IEEE Trans. Intell. Transp. Syst.* **2018**, *20*, 374–382.
55. Jiang, W.; Zhen, Y. A real-time EV charging scheduling for parking lots with PV system and energy store system. *IEEE Access* **2019**, *7*, 86184–86193.
56. Yilmaz, M.; Krein, P.T. Review of battery charger topologies, charging power levels, and infrastructure for plug-in electric and hybrid vehicles. *IEEE Trans. Power Electron.* **2012**, *28*, 2151–2169.
57. Darabi, Z.; Ferdowsi, M. Aggregated impact of plug-in hybrid electric vehicles on electricity demand profile. *IEEE Trans. Sustain. Energy* **2011**, *2*, 501–508.
58. Pasaoglu, G.; Fiorello, D.; Martino, A.; Scarcella, G.; Alemanno, A.; Zubaryeva, A.; Thiel, C. *Driving and Parking Patterns of European Car Drivers—A Mobility Survey*; European Commission Joint Research Centre: Luxembourg, 2012.
59. Qdr, Q. *Benefits of Demand Response in Electricity Markets and Recommendations for Achieving Them*; Tech. Rep. 2006; US Department Energy: Washington, DC, USA, 2006.
60. Moon, J.; Jung, T.Y. A critical review of Korea's long-term contract for renewable energy auctions: The relationship between the import price of liquefied natural gas and system marginal price. *Util. Policy* **2020**, *67*, 101132.
61. Kim, T.; Vecchietti, L.F.; Choi, K.; Lee, S.; Har, D. Machine Learning for Advanced Wireless Sensor Networks: A Review. *IEEE Sens. J.* **2021**, *21*, 12379–12397.
62. Shi, H.; Xu, M.; Li, R. Deep learning for household load forecasting—A novel pooling deep RNN. *IEEE Trans. Smart Grid* **2017**, *9*, 5271–5280.
63. Ni, P.; Li, Y.; Li, G.; Chang, V. Natural language understanding approaches based on joint task of intent detection and slot filling for IoT voice interaction. *Neural Comput. Appl.* **2020**, *32*, 16149–16166.
64. Borowy, B.S.; Salameh, Z.M. Optimum photovoltaic array size for a hybrid wind/PV system. *IEEE Trans. Energy Convers.* **1994**, *9*, 482–488.
65. Peterson, S.B.; Apt, J.; Whitacre, J. Lithium-ion battery cell degradation resulting from realistic vehicle and vehicle-to-grid utilization. *J. Power Sources* **2010**, *195*, 2385–2392.
66. Nengroo, S.H.; Ali, M.U.; Zafar, D.A.; Hussain, S.; Murtaza, T.; Alvi, M.J.; Raghavendra, K.V.G.; Kim, H.J. An Optimized Methodology for a Hybrid Photo-Voltaic and Energy Storage System Connected to a Low-Voltage Grid. *Electronics* **2019**, *8*, 176. https://doi.org/10.3390/electronics8020176.
67. Liu, C.-Y.; Zou, C.-M.; Wu, P. A task scheduling algorithm based on genetic algorithm and ant colony optimization in cloud computing. In Proceedings of the 2014 13th International Symposium on Distributed Computing and Applications to Business, Engineering and Science, Xi'an, China, 24–27 November 2014; pp. 68–72.
68. Ahmed, R.; Sreeram, V.; Mishra, Y.; Arif, M. A review and evaluation of the state-of-the-art in PV solar power forecasting: Techniques and optimization. *Renew. Sustain. Energy Rev.* **2020**, *124*, 109792.